\documentclass{ws-procs10x7}
\usepackage{balance, psfig}

\newcolumntype{d}[1]{D{.}{.}{#1}}

\def\@listI{\leftmargin\leftmargini
            \listparindent\parindent
            \parsep \z@\labelsep.5em
            \topsep 6.5\p@ \@plus3\p@
            \itemsep0\p@}

\makeindex
\begin{document}
\normalsize
\title{MEASUREMENTS OF TOP QUARK PROPERTIES AT CDF}

\author{A. C. Kraan$^*$}

\address{Department of Physics and Astronomy, University of Pennsylvania, Philadelphia PA-19104, USA.\\$^*$ Currently at: Istituto Nazionale di Fisica Nucleare, Sezione di Pisa, Pisa 56100, Italy.\\E-mail: Aafke.Kraan@cern.ch}


\twocolumn[\maketitle\abstract{The top quark with its mass of about 172 GeV/c$^2$ is the most massive fundamental particle observed by experiment.  In this talk we highlight the most recent measurements of several top quark properties performed with the CDF detector based on data samples corresponding to integrated luminosities up to 1~fb$^{-1}$. These results include a search for top quark pair production via new massive resonances, measurements of the helicity of the $W$ boson from top-quark decay, and a direct limit on the lifetime of the top quark.
}
]

\section{Introduction}
The top quark is the only elementary particle whose mass is of the same order of magnitude as the electroweak breaking scale, making it a highly interesting quark to study. Studying top-quark properties is important, not only because it might give us an inside look into the electroweak symmetry breaking within the Standard Model (SM), but also because unknown non-SM effects may enter, which could be absent in physics related to lighter quarks.  Although the top quark was discovered more than 10 years ago, relatively little is known about its properties, and measurements have so far been limited by statistics. The data which have recently been collected by CDF at Fermilab allows however for many improvements! 

Measurements of top-quark properties are based on the assumption that top quarks are strongly produced in pairs. In the SM, a top quark decays with almost 100\% into a $W$ boson and a b-quark. The $W$ boson decays either into two quarks, or into a charged lepton and a neutrino. Measurements are done using a 'lepton plus jets' or a 'dilepton' sample. The first is enriched in events where one of $W$ bosons decays leptonically and the other hadronically, \small{$p\bar{p}\rightarrow t\bar{t}\rightarrow W^+bW^-\bar{b}\rightarrow \bar{\ell}\nu b  q \bar{q'}\bar{b}$}\normalsize. The second sample is aimed on containing events where both $W$ bosons decay leptonically: \small{$p\bar{p}\rightarrow  t\bar{t}\rightarrow W^+bW^-\bar{b}\rightarrow \bar{\ell}\nu b \ell' \bar{\nu'}\bar{b}$}\normalsize. 

Here I discuss the most recent measurements of top-quark properties at CDF based on data samples up to 1 fb$^{-1}$. Measurements of the top-quark mass and pair production cross sections are discussed elsewhere\cite{florencia,chris}. 

\section{Resonanance search}
\begin{figure}[t!]
\hspace*{1cm}\psfig{file=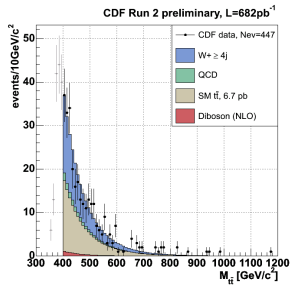,width=4cm}
\vspace{-1mm}
\caption{Distribution of the $t\bar{t}$ invariant mass.}
\label{res}
\vspace{-1mm}
\end{figure}

The production of top quark pairs via heavy resonances is predicted in several theoretical models such as technicolor. In CDF 682 pb$^{-1}$ of data have been analysed in order to search for resonances\cite{ttbarres}. The data sample used was the lepton plus jets sample with at least 4 jets. Candidate events are fully reconstructed and the $t\bar{t}$ invariant mass is studied. No evidence of resonance production is found (see Fig.~\ref{res}), and an upper limit is set on the production cross section of top quark pairs via resonances. In the framework of the so-called leptophobic topcolor model, resonances are excluded at 95\% confidence level (CL) with masses below 725 GeV/c$^2$. 

\section{W-helicity measurements}
\subsection{Introduction}
In the SM, the V-A weak interaction predicts a top quark with mass 175 GeV/$c^2$ to decay via $t\rightarrow W^+b$ with a fraction $f^0=70\%$ into longitudinally polarized $W$ bosons, and $f^-=30\%$ into left-handed polarized $W$ bosons.  Due to the small b-quark mass compared to the top-quark mass, the fraction of right-handed polarized $W$ bosons $f^+$ is less than $10^{-4}$. In contrast to the $W\ell\nu$ vertex, which V-A structure is experimentally accurately confirmed at LEP, the structure of the $tWb$ vertex is experimentally poorly explored. Measuring a deviation from the above values would indicate new physics. Past measurements have been strongly limited by statistics, and still leave plenty of space for new physics. The high luminosity data samples becoming available to CDF
allow for significant improvements now and in the near future.
\subsection{Sensitive variables}
 The structure of the $tWb$ vertex can be studied by measuring the helicity fractions of the $W$ boson.  The $W$-helicity is reflected in the angular distribution of its leptonic decay products. The small lifetime of the top quark ($<10^{-25}s$), which is much smaller than the typical hadronization time ($\sim 10^{-23}s$), assures that spin information is directly transferred to the $W$ boson and its decay products. Due to helicity conservation, a right-handed $W$ boson from top quark decay produces preferably a charged lepton opposite to the b-quark direction in the $W$ boson restframe, or in the direction of the $W$ boson itself when looking in the laboratory frame. For a left-handed $W$ boson the preferred direction is the opposite. The following discriminators can be used to measure the $W$-helicity, listed in order of complexity:

\begin{itemize}
\item
The \emph{transverse momentum of the charged lepton}\cite{leptonpt}. Because for a right-handed $W$ boson the direction of the charged lepton in the lab frame is the same as that of the $W$ itself, the lepton gets boosted. For a left-handed $W$ boson, the lepton gets anti-boosted. Thus, right-handed $W$'s produce higher $p_T$ charged leptons in the detector than left handed $W$'s. Longitudinal $W$'s have no preference (see third item).
\item The \emph{invariant mass squared $M_{lb}^2$} of the charged lepton and the b-quark from top-quark decay. Since the b-quark momentum is opposite to that of the $W$ boson when looking in the top-quark rest frame, including the b-quark dynamics provides additional sensitivity to the $W$-helicity, next to the lepton kinematics\cite{ourpaper}. The next item will clarify this. It should be noted that experimentally, there is an ambiguity when associating the b-quark and the charged lepton to the same top-quark.
\item  The \emph{cosine of the angle $\theta^*$}, where $\theta^*$ is the polar angle of the charged lepton in the $W$ boson restframe, with the z-axis defined to be the direction of the $W^+$-boson in the top-quark restframe. Due to helicity conserva-tion, charged leptons are produced according to:

\small
\vspace{0.2cm}
\begin{tabular}{lr}
\hspace*{-0.8cm}$d\sigma/d\cos\theta^*\sim (1-\cos\theta^*)^2$ & \hspace*{-0.3cm}(left-handed $W$)\\

\hspace*{-0.8cm}$d\sigma/d\cos\theta^*\sim (1+\cos\theta^*)^2$ &\hspace*{-0.3cm}(right-handed $W$)\\

\hspace*{-0.8cm}$d\sigma/d\cos\theta^*\sim (1-{\cos\theta^*}^2)$ &\hspace*{-0.3cm}(longitudinal $W$)
\end{tabular}
\normalsize
\\


\noindent It must be noted that theoretically $M_{lb}^2=1/2(M_t^2-M_W^2)(1+\cos\theta^*)$ in the limit $m_b\rightarrow 0$. Experimentally however, using $\cos\theta^*$ requires the full reconstruction and kinematic fit of the entire $t\bar{t}$ event\cite{shulamit,karlsruhe}, which is not necessary when using the $M_{lb}^2$ discriminator\cite{ourpaper}. On the other hand, the kinematic fit of the whole event resolves the ambiguity on the association of the charged lepton and the b-quark to the right top quark, and thereby increases the sensitivity. The $\cos\theta^*$ method can however only be applied to a subset of data (lepton plus at least 4 jets).
\end{itemize}

In CDF, recently three different analyses have been performed to measure the $W$ boson helicity based on the second and third observables above.
\normalsize
\subsection{$W$-helicity measurement based on $M_{lb}^2$}
The first approach is based on the $M_{lb}^2$ observable and uses 750 pb$^{-1}$ of data. Three independent data samples are used here. The first sample is a lepton plus jets sample with at least 3 jets of which one is b-tagged. The second sample is a lepton plus jets sample with at least 3 jets of which 2 are b-tagged. The third sample, not usable in the $\cos\theta^*$ method, is a dilepton sample without b-tagging requirements. A binned likelihood is used to extract $f^+$ while $f^0=07.$:

\vspace{2mm}
\footnotesize{
\hspace*{-0.5cm}$f^+\hspace{-0.7mm}=\hspace{-0.8mm}-0.02\hspace{-0.7mm}\pm \hspace{-0.7mm}0.07\hspace{0.5mm}\mathrm{(stat)}\hspace{0.5mm}\pm\hspace{-0.2mm} 0.04 \hspace{0.6mm}\mathrm{(syst)\hspace{0.9mm}}f^0\hspace{-0.1mm}=\hspace{-0.3mm}0.7 \mathrm{\hspace{1mm}fixed}$}
\vspace{2mm} 

\begin{figure}[t!]
\vspace{-0.2cm}
\psfig{file=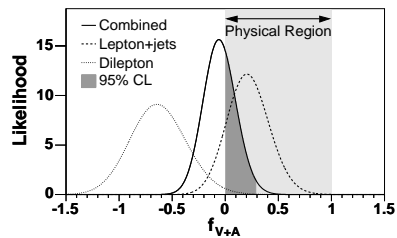,width=5.5cm}
\caption{The likelihood in data for the lepton plus jets and dilepton samples separately, and combined.}
\label{fig3}
\vspace{-0.2cm}
\end{figure}

\normalsize In Fig.~\ref{fig3} the likelihood distribution for the data is given. The lepton plus jets and the dilepton results are consistent at the 10\% confidence level.  The result for $f^+$ is consistent with the V-A prediction of the SM ($f^+=0$), and a 95\% CL upper limit on $f^+$ is set: $f^+<0.09$.

\subsection{$W$-helicity measurements based on cos$\theta^*$}
\normalsize{
The second approach\cite{shulamit} is based on studying the $\cos\theta^*$ distribution in $t\bar{t}$ candidate events using fully reconstructed lepton plus jets events, in which at least 4 reconstructed jets are required, of which one is tagged with a secondary vertex tagger. The analysis is based on 955 pb$^{-1}$ of CDF data. The kinematic fitter used for top-quark mass measurements is applied here to reconstruct the $t\bar{t}$ event, i.e. to find the best assignment of the top quark and its decay products to the detector objects, and to obtain the 4-vectors of the $W$, top and charged lepton. The assignment with the lowest fit $\chi^2$ is used to construct $\cos\theta^*$. The data events are compared with template distributions to determine the fractions $f^0$, $f^-$ and $f^+$. An unbinned likelihood method is used to extract the fractions in data:}

\footnotesize{
\hspace*{-0.5cm}$f^0\hspace{-0.3mm}=\hspace{1.5mm}0.61\hspace{-0.7mm}\pm \hspace{-0.7mm}0.12\mathrm{\hspace{0.8mm}(stat)\hspace{0.7mm}}\pm \hspace{-0.5mm}0.06\mathrm{\hspace{0.5mm}(syst)\hspace{2mm}}f^+\hspace{-0.7mm}=\hspace{-0.7mm}0 \mathrm{\hspace{1mm}fixed}$}

\vspace{2mm}
\footnotesize{
\hspace*{-0.5cm}$f^+\hspace{-0.7mm}=\hspace{-0.8mm}-0.05\hspace{-0.7mm}\pm \hspace{-0.7mm}0.06\hspace{0.5mm}\mathrm{(stat)}\hspace{0.5mm}\pm\hspace{-0.2mm} 0.03 \hspace{0.6mm}\mathrm{(syst)\hspace{0.9mm}}f^0\hspace{-0.1mm}=\hspace{-0.3mm}0.7 \mathrm{\hspace{1mm}fixed}$}
\vspace{2mm}

\begin{figure}[h!]
\vspace{-0.2cm}
\psfig{file=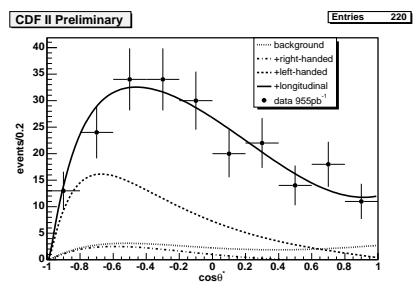,width=5.5cm}
\caption{The $\cos\theta^*$ distributions for data (points), overlayed with background (dotted), background + right-handed signal (dash-dot), background +right-handed +left-handed signal (dashed), and back-ground+right-handed+longitudinal (solid), when fitting for $f^+$ while $f^0=0.7$.}
\vspace{-0.3cm}
\label{fig1}
\end{figure}

\normalsize
In Fig~\ref{fig1} the observed $\cos\theta^*$ distribution is displayed when fitting for $f^+$ while $f^0=0.7$. A 95\% CL upper limit on $f^+$ is set: $f^+<0.11$. 

Another method making use of $\cos\theta^*$ is similar to the above method, however the reconstruction of $t\bar{t}$ events is done differently\cite{karlsruhe}. A different kinematic fitter is used, which also treats events with 5 jets, and which additionally uses $b$-jet probability\cite{jetprobpaper}. The data distribution is compared with template distributions and a binned likelihood is used to find the best fit value in data:
\vspace{2mm}

\footnotesize{
\hspace*{-0.5cm}$f^0\hspace{-0.3mm}=\hspace{1.5mm}0.59\hspace{-0.7mm}\pm \hspace{-0.7mm}0.12\mathrm{\hspace{0.8mm}(stat)\hspace{0.7mm}}\pm \hspace{-0.5mm}0.07\mathrm{\hspace{0.5mm}(syst)\hspace{2mm}}f^+\hspace{-0.7mm}=\hspace{-0.7mm}0 \mathrm{\hspace{1mm}fixed}$}

\vspace{2mm}
\footnotesize{
\hspace*{-0.5cm}$f^+\hspace{-0.7mm}=\hspace{-0.8mm}-0.03\hspace{-0.7mm}\pm \hspace{-0.7mm}0.06\hspace{0.5mm}\mathrm{(stat)}\hspace{0.5mm}\pm\hspace{-0.2mm} 0.04 \hspace{0.6mm}\mathrm{(syst)\hspace{0.9mm}}f^0\hspace{-0.1mm}=\hspace{-0.3mm}0.7 \mathrm{\hspace{1mm}fixed}$}
\vspace{2mm}

\begin{figure}[h!]
\vspace{-0.4cm}
\psfig{file=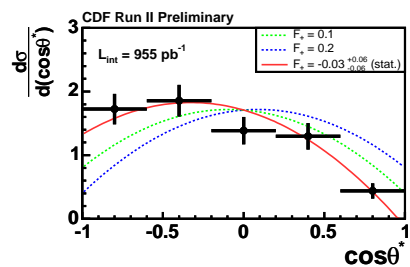,width=5.5cm}
\vspace{-0.2cm}
\caption{The $\cos\theta^*$ distributions for data (points), together with the best fit result (red) when fitting for $f^+$ while $f^0=0.7$. Also two dummy-fit values (blue, green) are displayed. }
\label{fig2}
\vspace{-0.4cm}
\end{figure}

\normalsize
The $\cos\theta^*$ distribution for data together with the fit result for $f^+$ while $f^0=0.7$ is displayed in Fig.~\ref{fig2}. Again the result is consistent with the SM, and an 95\% CL upper limit of $f^+<0.10$ is set.

\section{Top-quark lifetime}
\begin{figure}[t]
\psfig{file=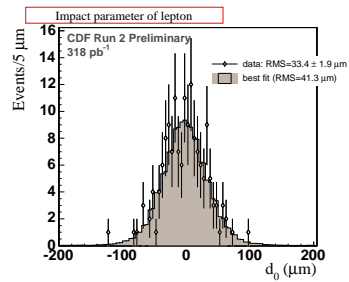,width=4.9cm}
\vspace{-1mm}
\caption{Distribution of $d_0$ for data and expected MC with $c\tau=0$.}
\label{life}
\end{figure}

In the SM, the lifetime of the top quark is predicted to be less than $\sim\hspace{-2mm}10^{-25}$ s, corresponding to a decay length of less than $\hspace{-2mm}\sim 10^{-10}\mu$m. Although the top-quark lifetime is theoretically strongly constrained by the CKM matrix, the experimental data leave plenty of space for a long lived top quark. In CDF the first direct measurement of the top-quark lifetime is done using 318 pb$^{-1}$ of Run~2 data\cite{lifetime}. A $t\bar{t}$ lepton plus jets sample is used with at least 3 jets of which one is b-tagged. The lifetime of the top quark is extracted from the impact parameter $d_0$ of the charged lepton with respect to the collision vertex. The resolution is determined by a $Z\rightarrow l^+l^-$ data sample. For a SM $t\bar{t}$ event, $d_0$ would be compatible with zero, and any measured $d_0$ deviating from this would imply new physics. The $d_0$ distribution for data is given in Fig.~\ref{life}. No deviation from the SM is seen, and a 95\% CL upper limit on the top-quark lifetime of $1.8*10^{-13}$ s or 53 $\mu$m is set.
\vspace{-1mm}
\section{Conclusions}
Information and updates on all top property measurements done at CDF are collected and available to the public\cite{all}. No deviation from the SM is observed. Most measurements are currently statistically limited. The excellent recent performance of the Tevatron, with 2 fb$^{-1}$ of integrated luminosity now delivered to CDF, promises further advances for measurements of the properties of the top quark and a sharper image of any effects from potential new physics.
\section*{Acknowledgments}
I thank Evelyn Thomson for her feedback on my presentation and on this manuscript.

\end{document}